# Theoretical investigation of the lattice thermal conductivities of II-IV-V$_2$ pnictide semiconductors

Victor Posligua,[a] Jose J. Plata,[a] Antonio M. Márquez,[a] Javier Fdez. Sanz,[a] Ricardo Grau-Crespo[b*]

[a]*Departamento de Química Física, Facultad de Química, Universidad de Sevilla, Seville 41012, Spain.*

[b]*Department of Chemistry, University of Reading, Whiteknights, Reading RG6 6DX, UK. Email: r.grau-crespo@reading.ac.uk*

## Abstract

Ternary pnictides semiconductors with II-IV-V$_2$ stoichiometry hold potential as cost-effective thermoelectric materials with suitable electronic transport properties, but their lattice thermal conductivities ($\kappa$) are typically too high. Gaining insight into their vibrational properties is therefore crucial to finding strategies to reduce $\kappa$ and achieve improved thermoelectric performance. We present a theoretical exploration of the lattice thermal conductivities for a set of pnictide semiconductors with ABX$_2$ composition (A = Zn, Cd; B = Si, Ge, Sn; and X = P, As), using machine-learning-based regression algorithms to extract force constants from a reduced number of density functional theory simulations, and then solving the Boltzmann transport equation for phonons. Our results align well available experimental data, decreasing the mean absolute error by ~3 Wm$^{-1}$K$^{-1}$ with respect to the best previous set of theoretical predictions. Zn-based ternary pnictides have, on average, more than double the thermal conductivity of the Cd-based compounds. Anisotropic behaviour increases with the mass difference between A and B cations, but while the nature of the anion does not affect the structural anisotropy, the thermal conductivity anisotropy is typically higher for arsenides than for phosphides. We identify compounds, like CdGeAs$_2$, for which nanostructuring to an affordable range of particle sizes could lead to $\kappa$ values low enough for thermoelectric applications.

# 1. Introduction

There are two major groups of ternary ABX$_2$ semiconductors with the chalcopyrite structure: ternary chalcogenides with I-III-VI$_2$ composition (A = Cu, Ag; B = Al, Ga, In, Tl; and X = S, Se, Te) and ternary pnictides with II-IV-V$_2$ composition (A = Zn, Cd; B = Si, Ge, Sn; and X = P, As, Sb). Pnictide chalcopyrites, like their chalcogenide counterparts, constitute a remarkably versatile material platform utilised in the field of semiconducting materials. These materials can be synthesised with cost-effectiveness, and their composition can be extensively adjusted due to their ability to accommodate different cations and anions.

The II-IV-V$_2$ ternary pnictides have been studied theoretically and experimentally for their potential applications in optoelectronic devices, including solar cells, due to their stability, dopability, high carrier mobility, and favourable optical absorption/emission properties [1-5]. Some of these properties make them also interesting materials for thermoelectric applications. Pnictide compounds often exhibit favourable electronic structures that enhance electrical transport, such as the coexistence of flat and highly dispersive bands near the Fermi level, which has motivated previous investigations of their thermoelectric properties [6-11].

In addition to the high mobility, dopability and favourable electronic properties, achieving high thermoelectric performance requires low thermal conductivity ($\kappa$). Due to the semiconducting nature of the materials, the electronic contribution to $\kappa$ is small and the lattice contribution is dominant (hereafter, we refer to the lattice thermal conductivity as $\kappa$, ignoring the electronic contribution). Unfortunately, ternary pnictide semiconductors with the chalcopyrite structure tend to have high $\kappa$ values in bulk form. For example, ZnBX$_2$ compounds, with B=Si, Ge, or Sn, and X=P or As, have been predicted to exhibit excellent electron transport properties [9]; however, measured bulk thermal conductivities are very high (*e.g.* up to ~35 Wm$^{-1}$K$^{-1}$ for single-crystal ZnGeP$_2$ at room temperature [12]). Still, variations of these compositions (*e.g.*, replacing Zn with Cd) and nanostructuring effects may lead to much lower values of $\kappa$, particularly at higher temperatures. Therefore, gaining a systematic understanding of phonon structure and transport in these pnictide semiconductors, as a function of composition and temperatures, and exploring the impact of nanostructuring on thermal conductivity becomes crucial for improving the potential of these materials for thermoelectric (as well as other) applications.

Given the sensitivity of $\kappa$ to the material's synthetic procedure, which influences grain size and defect chemistry, caution is needed when interpreting thermal conductivity trends between experimental measurements at different compositions conducted under dissimilar conditions. Computer modelling facilitates a direct comparison of intrinsic thermal conductivity behaviour across various compositions and temperatures. Unfortunately, accurately predicting the lattice thermal conductivity poses computational challenges. One of the most accurate approaches relies on solving the Boltzmann's transport equation (BTE) for phonons [13], which requires calculating second-order and third-order interatomic force constants (IFCs). Conventionally, these IFCs are determined by computing atomic forces in supercells for each symmetrically distinct displacement of atomic positions, using density functional theory (DFT) [1, 6, 14-18]. However, obtaining third-order IFCs in this way requires a considerable number of DFT calculations, making this step a bottleneck in the first-principle prediction of $\kappa$ [6, 14, 15, 17-20]. Recently, innovative algorithms have emerged that expedite the calculation of IFCs by leveraging machine learning and related techniques to enable the extraction of IFCs from a much smaller set of DFT calculations [21-24]. These advancements pave the way for the accurate calculation of $\kappa$ across a wide range of compositions, as we have shown before for chalcopyrite chalcogenides [25-27] and will demonstrate here for the chalcopyrite pnictides.

In this work, we have theoretically investigated the thermal conductivity of a range of pnictide semiconductors with compositions of $ABX_2$, where A = Zn or Cd, B = Si or Ge, and X = P or As. Additionally, two compositions with B = Sn ($CdSnAs_2$ and $CdSnP_2$) are included, as they are also known experimentally. We have excluded compositions for which a large degree of cation disorder is expected (*e.g.*, $ZnSnAs_2$ and $ZnSnP_2$) as they would require a theoretical framework beyond our current approach, to deal with the effect of alloy scattering of phonons. We will compare with experimental values whenever available [12, 28-36] to demonstrate the accuracy of our calculations, and then make systematic predictions for a comprehensive range of compositions, temperatures and particle sizes.

## 2. Methodology

### 2.1. DFT-based geometry optimisation and force evaluations

DFT calculations were conducted using the Vienna Ab Initio Simulation Package (VASP) code [37, 38], which uses a planewave expansion of the valence wavefunctions, together with the projector-augmented wave (PAW) method to account for core-valence interactions [38]. The number of valence electrons for each atom was determined based on the standards suggested by Calderon *et al*. [39]. Energies and forces were computed using the Perdew, Wang, Ernzerhof (PBE) generalised gradient approximation (GGA) functional [40], adding Grimme's D3 van der Waals corrections [41]. The kinetic energy cutoff of the plane-wave basis set expansion was set at 500 eV, which is 25% above the standard value for the chosen PAW potentials, to reduce Pulay stress errors. Equilibrium structures were found by energy minimisation until the forces on all atoms were less than $10^{-7}$ eV Å$^{-1}$ (the strict criterion for force convergence was needed for accurate phonon calculations). To minimise noise in the forces, an additional support grid was used for the evaluation of augmentation charges. Geometry optimisations were initially performed on the tetragonal conventional cells consisting of 16 atoms. The forces required to calculate the IFCs were then obtained using a 4 × 4 × 2 supercell (512 atoms), and in this case reciprocal space integrations were performed solely at the Γ point. We checked that increasing the grid density to a Γ-centred 2 × 2 × 2 mesh did not notably impact the results.

### 2.2. Force constant prediction and machine learning regression

The HiPhive package [22], based on machine-learning regression algorithms, enabled the extraction of second-, third-, and fourth-order force constants within optimised cutoff distances, from the DFT-calculated forces. Fourth-order force constants do not have a direct effect on the BTE model used here; however, their inclusion improved the regression for the force constant potential (FCP) model. Multilinear regression to the DFT forces was carried out to obtain the force constants, using the recursive feature elimination (RFE) algorithm, which efficiently selects the most relevant features for the regression [42]. The convergence of the FCP model parameters (number of distorted structures and cutoff distances) was tested by evaluating the variation in lattice thermal conductivities. Converged cutoff distances found for the $CuGaTe_2$ chalcopyrite in our previous work [25] (11, 6.2 and 4 Å for the second-, third-, and fourth-order force constants, respectively) were extrapolated here to the II-IV-$V_2$ chalcopyrite compositions, based on numbers of coordination shells (rather than absolute distances) for consistency. The number of distorted structures for DFT calculations was fixed to 18 for all compositions, which is enough for convergence (see Ref. [25]) and well below the >600 DFT calculations needed, for the same cutoff distances, in the traditional approach. To simplify the workflow, our wrapper code [43] was used in conjunction with the HiPhive program, automating the generation of distorted supercells, force calculations with VASP, and the construction of the machine-learned FCPs.

**2.3. Boltzmann's transport equation solution**

After constructing the FCP model, lattice thermal conductivities were determined by solving the BTE with the ShengBTE code [15]. We employed the full iterative procedure to go beyond the relaxation time approximation and computed scattering times, including isotopic and three-phonon scattering. To ensure accurate results, a Gaussian smearing of 0.1 eV and a dense mesh of 20 × 20 × 10 **q** points were used in all calculations, striking a balance between memory demand and the convergence of κ with the number of **q** points. In order to avoid the additional computational cost of computing Born effective charges, non-analytical contributions (NACs) were not considered, as tests in the I-III-$VI_2$ chalcopyrites showed that they had only a small effect on κ (<2.5%) [25], and the effect can be expected to be even smaller in the more covalent II-IV-$V_2$ chalcopyrites studied here. The scalar values reported represent one-third of the trace of the thermal conductivity tensor, and the anisotropic effects are discussed in detail. Throughout this study, the calculated lattice thermal conductivity will be compared with the total experimental values (lattice & electronic) due to expected negligible electronic contributions at low-mid temperatures.

## 3. Results and discussion

Accurate determination of cell parameters is crucial for obtaining reliable phonon properties. PBE-D3 geometry optimisations effectively reproduce the experimental lattice parameters of all compounds (**Figure 1**). As evidenced in our previous work with chalcopyrites [25], the uncorrected PBE functional overestimates the lattice parameters *a* and *c*, so the inclusion of D3 dispersion corrections allows us to obtain very low discrepancies with experimental values, of ~0.4% and ~0.8% in average for *a* and *c*, respectively. The PBE-D3 functional provides an excellent balance between computational cost and accuracy for phonon calculations, even without fully addressing the limitations of the GGA in predicting electronic structures.

In contrast with the case of the I-III-VI$_2$ chalcopyrites studied in Ref. [25], where the nature of the anion clearly had the strongest effect on the relative values of the cell parameters, for the II-IV-V$_2$ chalcopyrites the effect of the nature of the cations seems to be more significant. Interestingly, the trends are slightly different depending on the crystallographic direction: whereas for the *a* parameter the dominant effect is the nature of the A cation (Cd-based compounds having larger values than the Zn-based compounds), for the *c* parameter the effect of the A cation is not so pronounced, and the nature of the B cation seems to be at least as important (with the general trend Sn > Ge > Si). As a result, there is significant structural anisotropy, which will discuss in more detail later, in the context of the anisotropy of the thermal conductivity.

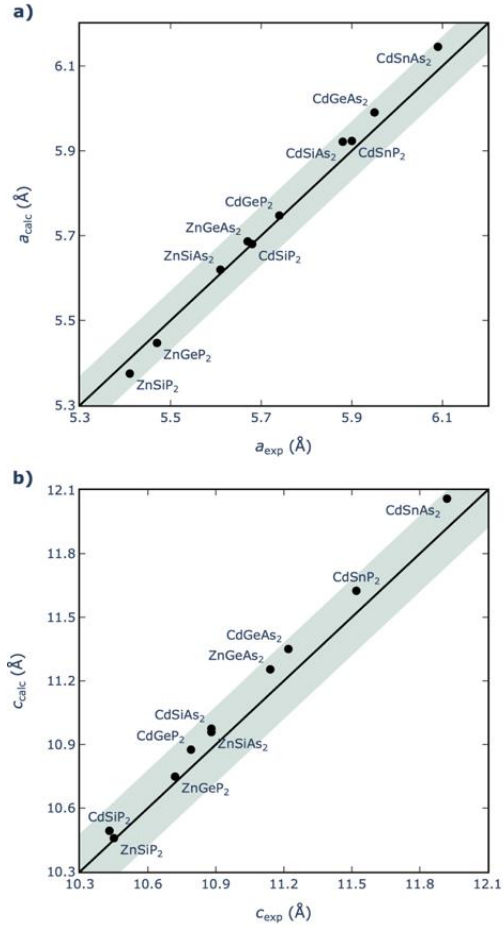

**Figure 1.** Comparison of experimental [44-52] and DFT-calculated values of a) *a* and b) *c* lattice parameters. Black solid line represents perfect agreement. Green-shaded area represents deviations of ≤0.8% from experiments in either direction.

**Table 1.** Lattice thermal conductivities (κ) for all the investigated ternary pnictide compounds, at 300 and 700 K.

| Compound | κ (W m$^{-1}$ K$^{-1}$) | |
|---|---|---|
| | 300 K | 700 K |
| CdGeAs$_2$ | 6.6 | 2.8 |
| CdGeP$_2$ | 15.9 | 6.7 |
| CdSiAs$_2$ | 7.4 | 3.2 |
| CdSiP$_2$ | 13.6 | 5.5 |
| CdSnAs$_2$ | 8.1 | 3.5 |
| CdSnP$_2$ | 21.2 | 9.2 |
| ZnGeAs$_2$ | 17.9 | 7.5 |
| ZnGeP$_2$ | 46.5 | 19.9 |
| ZnSiAs$_2$ | 20.4 | 8.6 |
| ZnSiP$_2$ | 29.7 | 11.8 |

The calculated isotropic averages of the thermal conductivities for all pnictide compositions are summarised in **Table 1** at 300 K and 700 K. An evident pattern is noticed when comparing Cd- and Zn-based pnictides: those based on Cd exhibit lower κ compared to their counterparts based on Zn. Across the compositions involving B = Si, Ge, Sn, and X = As, P, the average value of κ for Zn-based pnictides (27.3 Wm$^{-1}$K$^{-1}$) is approximately twice that of Cd-based compositions (12.1 Wm$^{-1}$K$^{-1}$). To investigate the origin of this behaviour, phonon density of states (pDOS) plots have been generated in selected compounds (CdGeAs$_2$ and ZnGeAs$_2$) to identify the contributions stemming from the A$^{2+}$ cations (**Figure 2**). Evidently, in the case of Cd, there are contributions to modes with frequencies below 1 THz, something which is not observed in the Zn-based composition. Despite both compounds having a similar distribution of group velocities (see **Figure 3a**), the introduction of low-frequency optical modes by Cd$^{2+}$ cation results in high scattering rates ($W_{ahn}$) at these frequencies. These scattering rates predominantly influence the thermal conductivity behaviour, as exhibited in **Figure 3b**.

The nature of the pnictogen X$^{3-}$ anion also affects the thermal conductivity, with the κ values of the arsenides being generally lower than that of their phosphide counterparts. On the other hand, the influence of the nature of the B cation is less predictable. For instance, B = Ge is found in both extremes of the distribution of thermal conductivities among all pnictide compositions in this study: CdGeAs$_2$ (6.6 W m$^{-1}$ K$^{-1}$) and ZnGeP$_2$ (46.5 W m$^{-1}$ K$^{-1}$) have the lowest and the highest κ values, respectively. This does not imply that κ is unaffected by the nature of B$^{4+}$ cations, but simply that that trends cannot be generalised in the same manner as observed for the ions within A and X sites, similarly to what we reported in our previous study of lattice thermal conductivities of chalcogenide chalcopyrites [25].

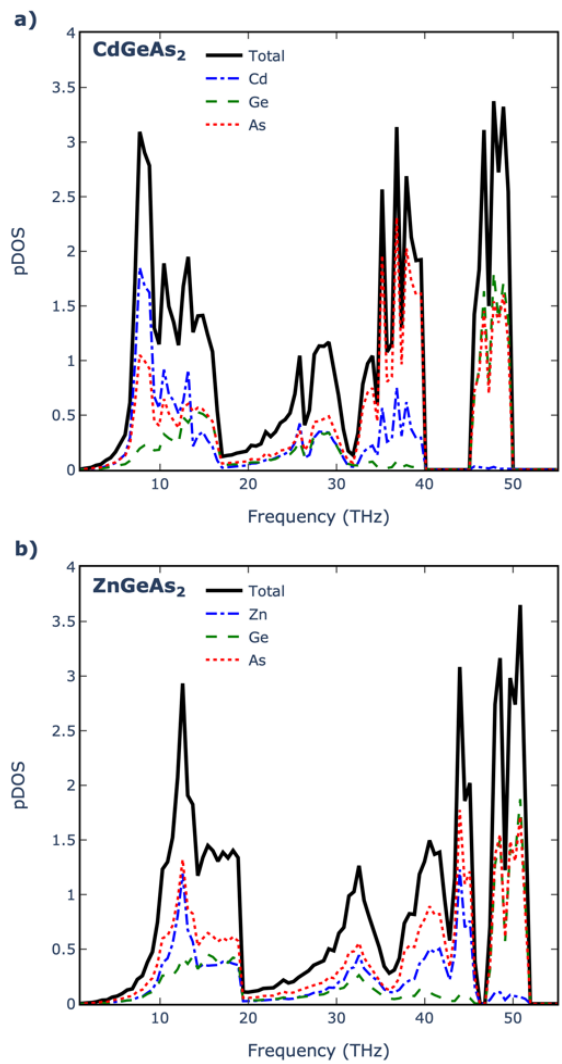

Figure 2. Phonon density of states (pDOS) for a) $CdGeAs_2$ and b) $CdGeAs_2$ pnictides.

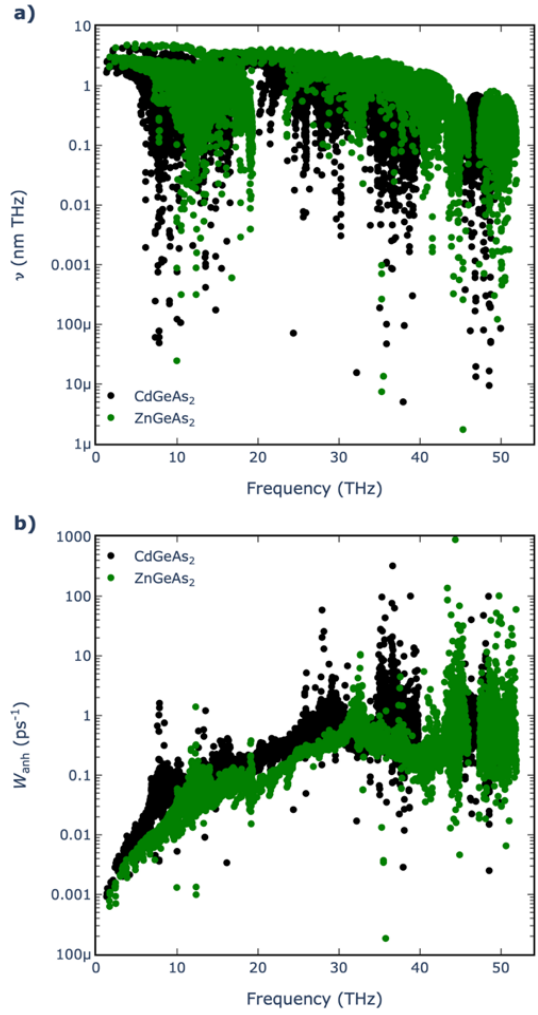

**Figure 3.** a) Group velocities and b) scattering rates vs mode frequency for CdGeAs$_2$ and ZnGeAs$_2$ pnictides.

Having examined the trends in the isotropic average values of κ, we now focus on the anisotropic behaviour. We distinguish here between the structural anisotropy, characterised by the deviation of the *c/a* ratio from the value of 2, and the thermal conductivity anisotropy, characterised by the deviation of the κ$_z$/κ$_x$ ratio from 1. Since the origin of the anisotropic behaviour in this family of materials is the distinction between the A and B cations (if the two cations were the same, we would recover the isotropic, cubic zincblende structure), it makes sense to study the variation of both *c/a* and κ$_z$/κ$_x$ versus the atomic mass difference ($\Delta M_{A-B}$) between the cations, as we have done in **Figure 4**. As a general trend, the larger the difference in mass between A and B, the more anisotropy (both structural and thermal) there is in the system.

Consistently with the discussion above about the cell parameters, the structural anisotropy is not significantly affected by the nature of the anions. For example, compounds with A = Cd and B = Si exhibit the strongest deviation from $c/a = 2$, followed by the Cd-Ge combinations. However, the thermal conductivity anisotropy is clearly affected by the nature of the anion: arsenides generally have stronger deviations from $\kappa_z/\kappa_x = 1$ than phosphides, which can be expected due to the higher covalence of the former compared to the latter. As a result, the most anisotropic thermal conductivity is found for CdSiAs$_2$, which has $\kappa_z/\kappa_x = 0.796$ at room temperature.

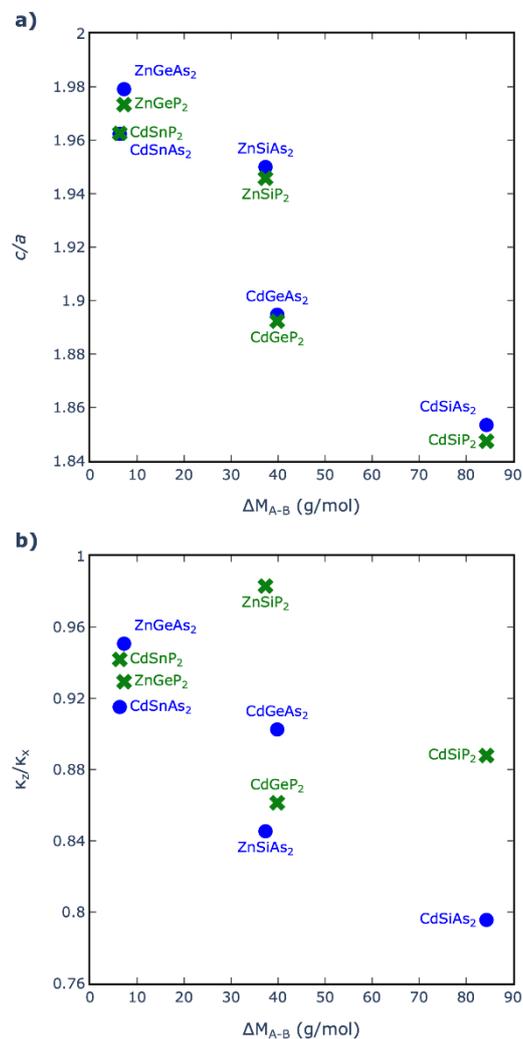

**Figure 4.** Variation of a) the *c/a* ratio (characterising the structural anisotropy) and of b) the $\kappa_z/\kappa_x$ ratio (characterising the thermal conductivity anisotropy) with the absolute difference in molar mass between the two cations A and B.

**Figure 5** presents an overview of the comparison between our findings, along with previous theoretical outcomes, and experimental measurements of κ conducted at room temperature. Our calculations in this study exhibit the closest alignment with experimental data within these range of pnictide compositions (**Figure 5**a). This agreement is to be expected considering that we are using a more advanced model for κ evaluation in comparison with prior studies. For instance, Toher *et al.* [53] adopted a more approximate technique that involved combining the Slack equation [54] with the Debye temperature and Grüneisen parameter, both extracted from DFT calculations using a quasi-harmonic Debye model. Although this technique operates independently of experimental parameters and its computational efficiency renders it suitable for a wide array of materials, it notably underestimates the thermal conductivity of all pnictide compositions. Yan *et al.* [35] introduced a methodology grounded in the Debye-Callaway model [55], exhibiting reasonable accuracy within a single order of magnitude across a large experimental dataset. While this approach led to improved results compared to those based on the Slack equation, a parameter fitting based on experimental data is required. Nevertheless, this method also tends to underestimate the κ values for all pnictides, as in Toher's method. In contrast, the approach used in our study enhances the calculation accuracy, boasting a mean absolute error (MAE) lower (by ~3 $Wm^{-1}K^{-1}$) than the best previous theoretical work (see **Figure 5**b). This is achieved without relying on experimental input and at notably reduced computational cost when contrasted with the conventional DFT-based method for obtaining the force constants to solve Boltzmann's transport equation.

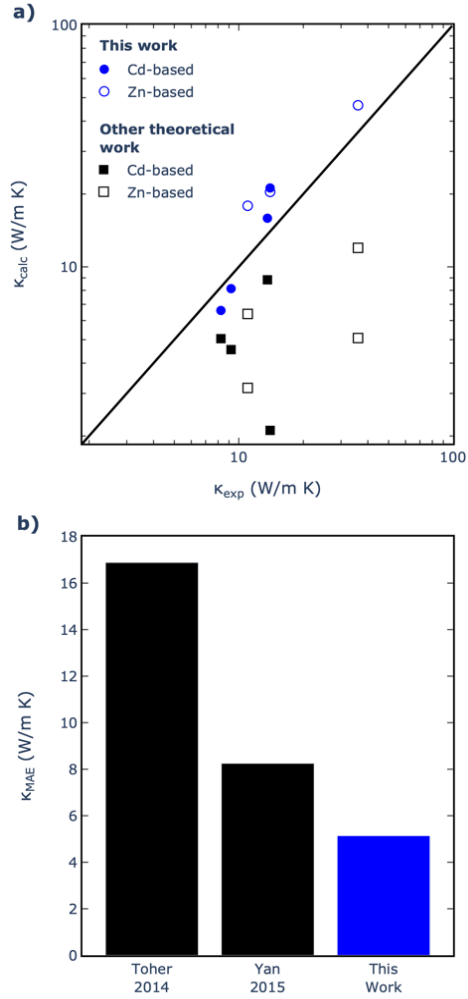

**Figure 5.** a) Comparison of room-temperature (300 K) κ values obtained from calculations in this work (blue circles) and previous theoretical work (black squares) with available experimental data. For both cases, solid and empty symbols denote Cd-based and Zn-based pnictides, respectively. Black line represents perfect agreement with experiment. b) Mean absolute error (MAE) against experimental values in this study relative to those observed in earlier theoretical analyses. Toher 2014 is Ref. [53] and Yan 2015 is Ref. [35]. Experimental data from Refs. [12, 28-36].

We can also contrast the projected temperature-dependent changes in κ with available experimental data from Zhang *et al.* [31] on CdSiP$_2$. **Figure 6** depicts the characteristic variation of thermal conductivities as a function of temperature *T*. The thermal conductivity primarily stems from phonon-phonon Umklapp scattering, consequently yielding a $T^{-1}$ variation. This relationship reflects the growing number of phonons that contribute to scattering as temperature rises.

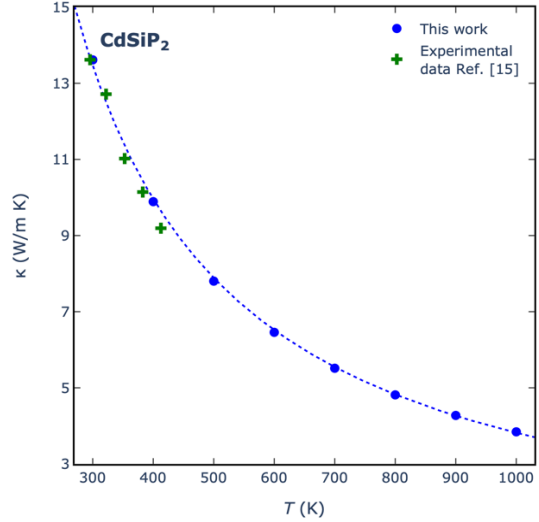

**Figure 6. Calculated temperature variation (blue) of κ for CdSiP$_2$ in comparison with the experimental data (green) reported in Ref. [31].**

Lastly, we explore the impact of nanostructuring on thermal conductivities using an approach that separates the contributions to κ by the phonon mean free path [56]. This method, widely employed in theoretical investigations of how nanostructuring affects thermal transport in thermoelectric materials [57-61], estimates the κ value associated with a specific particle size $L$ by considering the cumulative contributions from all mean free paths up to $L$ (i.e. subtracting contributions from mean free path exceeding the particle size). **Figure 7** shows an example, for CdSiP$_2$, of how the thermal conductivity is expected to vary with particle size. It clearly illustrates that nanostructuring at the micrometre scale already exert a perceptible influence on the thermal conductivity of this compound.

To quantitatively assess the trend of κ reduction due to nanostructuring across all compositions, we have calculated the particle size ($L_{0.5}$) that results in a 50% reduction from the bulk value, which is a function of temperature. **Figure 8** shows that the lower the bulk value of κ, the smaller the $L_{0.5}$. The particle size required to reduce the thermal conductivity by half is around 20 nm per each Wm$^{-1}$K$^{-1}$ of the bulk value of κ. This matches the trend observed for I-III-VI$_2$ chalcopyrites in Ref. [25], but here the correlation between $L_{0.5}$ and the bulk κ is not as strong.

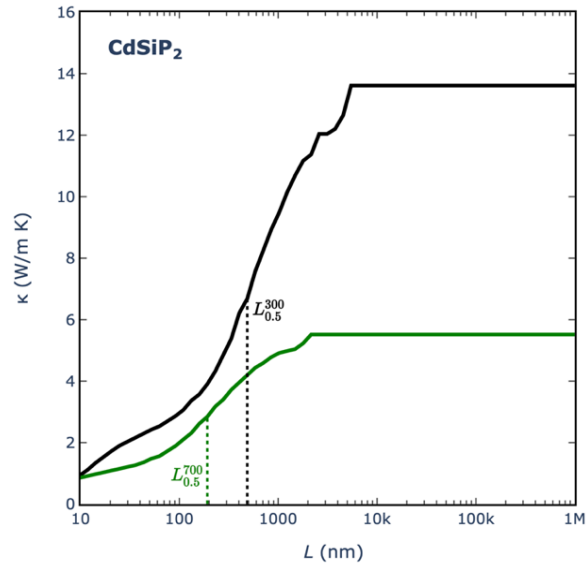

**Figure 7.** Cumulative lattice thermal conductivity from mean-free-path contributions up to distance $L$ for CdSiP$_2$, illustrating the effect that nanostructuring would have on their thermal conductivity. Black and green lines denote $T$ = 300 and 700 K, respectively.

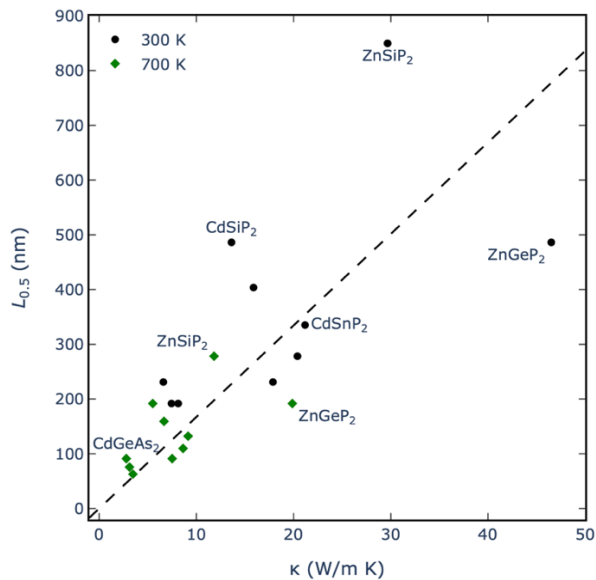

**Figure 8.** Correlation between $L_{0.5}$ and (bulk) $\kappa$, including points at 300 K (black circles) and 700 K (green diamonds). The dashed line corresponds to a proportionality constant of 20 nm/W m$^{-1}$ K$^{-1}$ between $L_{0.5}$ and $\kappa$.

The fact that the correlation between $L_{0.5}$ and bulk $\kappa$ is not perfect, suggests a strategy to identify compounds within this family that are particularly susceptible to reduction of thermal conductivity by nanostructuring. Compounds for which the points in **Figure 8** lie above the linear regression line require less particle size reduction than average to achieve a lower κ. For example, the room-temperature thermal conductivity of CdSiP$_2$ can be halved with respect to its bulk value, with particle sizes just around 500 nm. However, just halving the value of κ is not useful for thermoelectric applications of this compound, because its thermal conductivity at room temperature remains too high (according to **Figure 7**, we would need to go to particle sizes of around 10 nm to achieve a κ of ~1 Wm$^{-1}$K$^{-1}$ for this compound). For thermoelectric applications, it is more interesting to look at compounds for which the bulk κ is already low to start with, while at the same being more susceptible than average to κ reduction via nanostructuring.

At 700 K, CdGeAs$_2$ is a case for which a particle size reduction to around the affordable value of 100 nm, halves the thermal conductivity, to ~1.4 Wm$^{-1}$K$^{-1}$. A recent theoretical study [62] has suggested that this material exhibits suitable electron transport properties for thermoelectric applications, for example, a high Seebeck coefficient above 500 μV K$^{-1}$ at 600 K when p-doped at a concentration of 10$^{18}$ cm$^{-3}$, and moderate electrical conductivity. Using the constant relaxation time approximation with τ=10$^{-14}$ s, these authors estimated thermoelectric figures of merit $zT$ of 0.26 at 600 K for the p-type compound at that level of doping, assuming a thermal conductivity of 4 Wm$^{-1}$K$^{-1}$. Our results here show that a much lower thermal conductivity than that, and therefore a proportionally higher $zT$, can be achieved for this compound via nanostructuring to an affordable particle size of ~100 nm.

## 4. Conclusions

We have performed a systematic investigation of the lattice thermal conductivities of ternary pnictide chalcopyrites, using a combination of Boltzmann transport theory, density functional theory simulations, and machine-learning regression algorithms to calculate force constants. We have gained significant insights into the factors governing the thermal conductivity behaviour of these II-IV-V$_2$ semiconductors upon variations in chemical composition, temperature, and nanostructure.

Our study has demonstrated a clear contrast in thermal conductivity between Cd- and Zn-based pnictides. This divergence predominantly stems from the lower frequencies associated with vibrational modes involving Cd atoms. These vibrational modes exhibit an overlap with acoustic modes, leading to an amplified scattering process and reduced scattering times. While clear trends are elusive when substituting the B cation, the behaviour of the pnictogen X atom follows a non-monotonic pattern down the group. Particularly, arsenides exhibit lower thermal conductivities than their corresponding phosphides.

An evident but moderate κ anisotropy is observed, with Cd-based pnictides showing greater anisotropy than their Zn-based counterparts. This alignment echoes the structural anisotropy indicated by the *c/a* ratio. Accurate predictions of room-temperature lattice thermal conductivities and the correspondence between κ and temperature trends with experimental data corroborates the reliability of our findings.

Lastly, we have delved into the influence of grain size on κ, by examining the particle sizes required to halve the thermal conductivity from the bulk values. We have identified interesting cases, like CdGeAs$_2$, for which moderate particle size reduction can lead to thermal conductivities sufficiently low for thermoelectric applications. This is important because several II-IV-V$_2$ compounds have been shown to exhibit suitable electron transport properties for thermoelectric applications but generally suffer from too high thermal conductivities. We hope that our work will motivate further investigation of the thermoelectric potential of this family of compounds.

## Acknowledgements

This work was funded by MICIN/AEI/10.13039/501100011033 and by "European Union Next Generation EU/PRTR" (grants PID2019-106871GB-I00 and TED2021-130874B-I00, PID2022-

138063OB-I00). We thankfully acknowledge the computer resources at Lusitania and the technical support provided by Cénits-COMPUTAEX and Red Española de Supercomputación, RES (QS-2022-2-0030, QHS-2022-3-0032). This work also made use of the Young supercomputer, via the UK's Materials and Molecular Modelling Hub, which is partially funded by EPSRC (EP/T022213/1 and EP/W032260/1).